%% file: main.tex
\documentclass[cameraready]{Interspeech}

\usepackage{amssymb}
\usepackage{pifont}

\definecolor{almondlow}{RGB}{252,239,219} 
\definecolor{almondmiddle}{RGB}{237,225,206} 
\definecolor{almondhigh}{RGB}{224,213,194} 
\definecolor{almondultra}{RGB}{214,204,186} 
\definecolor{melon}{RGB}{232,156,124} 
\definecolor{champagne}{rgb}{0.97, 0.91, 0.81}
\definecolor{baselinecolor}{rgb}{1,1,1}
\definecolor{corn}{rgb}{0.98, 0.93, 0.36}
\definecolor{teagreen}{rgb}{0.82, 0.94, 0.75}
\definecolor{predcolor}{rgb}{0.74, 0.83, 0.9}
\definecolor{decorrcolor}{rgb}{0.98, 0.85, 0.87} 
\definecolor{camouflagegreen}{rgb}{0.47, 0.53, 0.42}
\definecolor{brightube}{rgb}{0.82, 0.62, 0.91} 
\definecolor{paletaupe}{rgb}{0.74, 0.6, 0.49}
\definecolor{pastelviolet}{rgb}{0.8, 0.7, 0.79}
\definecolor{knowcolor}{rgb}{0.80, 0.94, 0.75}
\definecolor{offlinecolor}{rgb}{1,1,1}
\definecolor{firsttaskcolor}{rgb}{0.97, 0.97, 0.97}
\definecolor{azure(colorwheel)}{rgb}{0.0, 0.5, 1.0}
\definecolor{gray(x11gray)}{rgb}{0.75, 0.75, 0.75}
\definecolor{lightgray}{rgb}{0.90, 0.90, 0.90}
\definecolor{darkgray}{rgb}{0.66, 0.66, 0.66}
\definecolor{pastelorange}{rgb}{1.0, 0.7, 0.28}
\definecolor{pastelyellow}{rgb}{0.99, 0.99, 0.59}
\definecolor{pistachio}{rgb}{0.58, 0.77, 0.45}
\definecolor{shamrockgreen}{rgb}{0.0, 0.62, 0.38}
\definecolor{pastelred}{rgb}{1.0, 0.41, 0.38}
\definecolor{slateblue}{rgb}{0.32, 0.55, 0.8}
\definecolor{pinksecondbest}{RGB}{252, 241, 241}
\definecolor{aqua}{rgb}{0.0, 1.0, 1.0}
\definecolor{aquamarine}{rgb}{0.5, 1.0, 0.83}

\definecolor{pastelred}{RGB}{232, 131, 131}
\definecolor{pastelviolet}{rgb}{0.8, 0.75, 0.92}
\definecolor{champagne}{rgb}{0.97, 0.91, 0.81}
\definecolor{lightblue}{RGB}{225, 241, 255}
\usepackage{soul}

\usepackage{tcolorbox}

\usepackage{amsmath,graphicx,hyperref}

\usepackage{xcolor}

\usepackage{url}
\usepackage{comment}
\usepackage{booktabs}
\usepackage{xcolor}
\usepackage{colortbl}
\usepackage{subcaption}
\usepackage{arydshln}
\usepackage{multirow}
\usepackage{pifont}
\usepackage{hyperref}
\usepackage{enumitem}
\usepackage{cite}
\usepackage{soul}
\usepackage{xcolor}

\title{MambAdapter: Lightweight Mamba-Based Adapters for Parameter-Efficient Transfer Learning in Speech and Audio}

\author[affiliation={1,4}]{Salman}{Hussain Ali}
\author[affiliation={2}, orcid=0009-0006-3443-5143]{Umberto}{Cappellazzo}
\author[affiliation={3,4}]{Mirco}{Ravanelli}

\address{%
    $^{1}$Université de Montréal, Canada \quad
    $^{2}$Imperial College London, UK \quad
    $^{3}$Concordia University, Canada \quad
    $^{4}$Mila -- Quebec AI Institute, Canada%
}

\email{salman.sami.hussain.ali@umontreal.ca, mirco.ravanelli@concordia.ca}


\usepackage{comment}

\begin{document}

\maketitle

\input{Sections/Abstract}

\input{Sections/Introduction}

\input{Sections/Related_Work}

\input{Sections/Methodology}

\input{Sections/Experiments}

\input{Sections/Conclusion}

\section{Acknowledgments}
We gratefully acknowledge the support of NSERC, the Digital Research Alliance of Canada (alliancecan.ca), Translated (Imminent Program), and Apple (Seed Grant) through research funding, computing resources, and donations.

\section{Generative AI Use Disclosure}
Generative AI tools were used during the preparation of this paper to improve clarity and grammar, as well as to assist with coding tasks such as refactoring and debugging. No AI tools were used to generate the scientific ideas, experimental design, results, or analysis presented in this work. All content has been reviewed and verified by the authors, who assume full responsibility for the accuracy and integrity of the paper.

\bibliographystyle{IEEEtran}
\bibliography{refs}

\end{document}

%% file: Sections/Abstract.tex
\begin{abstract}
\label{sec:Abstract}
Fine-tuning Transformer-based foundation models has become the dominant strategy for domain adaptation in audio and speech processing. To reduce the computational and memory costs of this process, parameter-efficient transfer learning (PETL) methods have been widely explored. Meanwhile, \textit{Mamba}, a recent state-space model, has emerged as a promising alternative to Transformers for sequence modeling. In this work, we present \textbf{MambAdapter}, a parameter-efficient transfer learning approach that integrates Mamba into low-rank bottleneck adapters. Our design combines parameter sharing across adapters with the injection of a lightweight Mamba module, enabling more effective modeling of audio features. We demonstrate that \textbf{MambAdapter} matches or outperforms strong PETL baselines on four audio classification tasks and five speech recognition languages, even when operating under reduced parameter budgets.    
\end{abstract}

%% file: Sections/Introduction.tex
\section{Introduction}
\label{sec:introduction}

Recent advances in speech and audio processing have been largely driven by the emergence of large-scale foundation models. These models, such as Whisper~\cite{radford2022robustspeechrecognitionlargescale}, are pre-trained on massive multilingual datasets and demonstrate strong zero-shot and few-shot capabilities across a variety of downstream tasks. Directly fine-tuning these models on downstream tasks is a highly expensive and cumbersome process. So, parameter-efficient transfer learning (PETL) techniques emerged as the primary approach to adapting these models with lower cost and higher efficiency~\cite{houlsby2019parameterefficienttransferlearningnlp, pfeiffer-etal-2021-adapterfusion, hu2021loralowrankadaptationlarge}. These techniques have been applied in the context of speech and audio processing as well~\cite{cappellazzo2024parameterefficienttransferlearningaudio, cappellazzo2024efficient, 10096311, thomas2022efficient, kessler2022adapter}, and strong empirical results support their use.

In parallel, prominent developments in the field of sequence modeling include Mamba\cite{gu2024mamba}, a selective state space model, as an alternative to the Transformer\cite{vaswani2023attentionneed} for signal processing, boasting linear-time sequence modeling while maintaining competitive performance with Transformers. Mamba has shown strong performance in various applications in the speech domain, from speech separation\cite{10888514}, self-supervised learning\cite{335de060a3b74de3b791d8f7c318ce88}, and speech recognition\cite{fang2024mambauma ,Zhang2024MambaIS}.  It is particularly effective for speech because its linear-time selective state-space design efficiently models both short- and long-range patterns, with the added benefit of lower latency than Transformers. However, its use in the context of parameter-efficient transfer learning has not yet been explored. Utilizing Mamba as a PETL technique in speech would enable the model to efficiently capture patterns in the long-context inherent in speech at a relatively low computational cost, while taking advantage of Mamba's strength in modeling long-context sequences.

As such, we investigate the following question:
\begin{tcolorbox}[colback=blue!5,colframe=blue!20,arc=2mm,outer arc=1mm]
\textbf{(}$\mathbf{Q}$\textbf{)} \textit{Can Mamba be leveraged as a parameter-efficient transfer learning technique for audio and speech?}
\end{tcolorbox}

To address this question, we explore an approach to PETL that involves using Mamba within lightweight bottleneck adapters\cite{houlsby2019parameterefficienttransferlearningnlp}. We introduce \textbf{MambAdapter}, a highly parameter-efficient transfer learning technique that uses shared linear projections and low-rank Mamba blocks~\cite{gu2024mamba}. This approach enables us to leverage Mamba’s strong sequential modeling capabilities together with the frozen Transformer backbone, preserving its strengths while enabling efficient adaptation. We use MambAdapter to adapt the Audio Spectrogram Transformer (AST)\cite{gong2021astaudiospectrogramtransformer} on four audio and speech classification datasets, and Whisper\cite{radford2022robustspeechrecognitionlargescale} on five different languages for multilingual speech recognition. \textit{Our results show that MambAdapter matches or outperforms existing PETL methods while using significantly fewer trainable parameters.} We also analyze how MambAdapter scales with parameter budget and conduct ablation studies to assess the contribution of each architectural component. Additionally, we demonstrate the effectiveness of parameter sharing in adapters and analyze the performance-efficiency
trade-off. To the best of our knowledge, this is the first work to incorporate Mamba as a parameter-efficient transfer learning technique into transformers.

\begin{figure}
    \centering
    \includegraphics[width=1.0\linewidth]{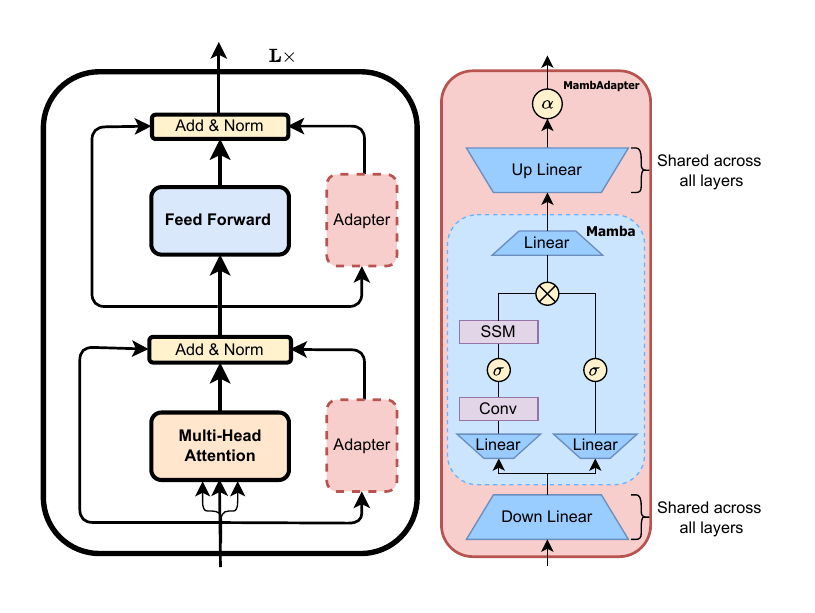}
    \caption{\textbf{Left:} A Transformer block with inserted adapter modules (dashed boxes). \textbf{Right:} The internal structure of MambAdapter, our proposed adapter.}
    \label{fig:MambAdapter-architecture}
\end{figure}

%% file: Sections/Related_Work.tex
\section{Related Work}
\label{sec:related-work}

\subsection{Overview of the Main PETL Methods}

\textbf{Bottleneck} adapters\cite{houlsby2019parameterefficienttransferlearningnlp, pfeiffer-etal-2021-adapterfusion} enable parameter-efficient transfer learning by inserting small task-specific modules with a bottleneck architecture into each Transformer layer, while keeping the original model's parameters frozen. \textbf{Conformer} adapters \cite{cappellazzo2024parameterefficienttransferlearningaudio} extend bottleneck adapters for speech tasks. They leverage the convolutional module of the Conformer model \cite{gulati2020conformerconvolutionaugmentedtransformerspeech} and introduce a lightweight convolutional block, where the input is processed using a sequence of depthwise and pointwise convolutions. Two main adapter insertion strategies exist: in the \textit{Pfeiffer}\cite{pfeiffer-etal-2021-adapterfusion} configuration, adapters are placed solely after the feed-forward network (FFN), whereas in the \textit{Houlsby}\cite{houlsby2019parameterefficienttransferlearningnlp, karimi-mahabadi-etal-2021-parameter} configuration they are inserted after both the attention and the FFN.

\textbf{LoRA}\cite{hu2021loralowrankadaptationlarge} performs parameter-efficient transfer learning by decomposing task-specific updates for pretrained weight matrices into low-rank factors. In practice, LoRA modules are commonly applied to the query and value projection matrices of each multi-head self-attention layer.

\subsection{Mamba}

\textbf{Mamba}~\cite{gu2024mamba} is a linear-time state-space model (SSM) that replaces self-attention with a selective state-space recurrence. It is derived from a continuous-time linear state-space system:
\begin{equation}
\label{eq:mamba-ct}
h'(t) = A h(t) + B x(t), 
\qquad 
y(t) = C h(t) + D x(t),
\end{equation}
where $A \in \mathbb{R}^{N \times N}$ defines the state transition dynamics, $B \in \mathbb{R}^{N \times 1}$ controls how the input drives the latent state, $C \in \mathbb{R}^{1 \times N}$ represents the output projection, and $D$ is the skip connection. The state dimension $N$ ( also denoted $d_{\text{state}}$) determines the capacity of the latent dynamical system.

For discrete sequences, the system is discretized with step size $\Delta$, yielding
\begin{equation}
h_t = \bar{A} h_{t-1} + \bar{B} x_t, 
\qquad 
y_t = C h_t + D x_t.
\end{equation}
Although expressed recurrently, this formulation is equivalent to a structured convolution with a kernel derived from the state-space parameters. In practice, Mamba augments this SSM core with a learnable local convolution of kernel size $k$, which captures short-range interactions before the state update.

Mamba further introduces selectivity, making the state transition parameters input-dependent via lightweight gating mechanisms. An expansion factor (\textit{expand}) controls the width of intermediate representations surrounding the SSM block, thereby scaling the model’s expressive capacity while preserving linear-time $\mathcal{O}(Nd_{state})$ complexity.

%% file: Sections/Methodology.tex
\section{MambAdapter}
\label{sec:methodology}

Figure \ref{fig:MambAdapter-architecture} presents the architecture of MambAdapter, our proposed adapter module designed for speech and audio tasks. MambAdapter aims to achieve both high parameter efficiency and strong performance on temporally complex inputs. It does so by combining a shared-projection bottleneck design with Mamba blocks, enabling the model to capture long-range temporal patterns within the low-rank latent space. 

To improve the performance of standard bottleneck adapters, we introduce a Mamba block and a learnable scaling factor $\alpha$ in each adapter layer. The Mamba block models the structured, long-range dependencies typical of audio and speech. Its implicit recurrence and large convolutional kernel make it particularly well-suited for processing continuous signals spanning long contexts. Also, by operating within the low-rank latent space, the Mamba block captures temporal dynamics with minimal overhead. It can be formulated as such:
\begin{equation}
X_{out}=\alpha \cdot Mamba(\hat X W_{down})W_{up} + F({\hat X})
\end{equation}
where $\alpha$ is the learnable scaling factor (initialized to 0.1), $\hat X$ is the input, $W_{\text{down}} \in \mathbb{R}^{d \times r}$ and $W_{\text{up}} \in \mathbb{R}^{r \times d}$ are the shared down- and up-projections with bottleneck rank $r \ll d$, and F is the underlying Transformer module (FFN or Attention).

While inserting Mamba blocks improves the representational capacity of each adapter, it also increases the number of trainable parameters. Specifically, each Mamba block contributes roughly $3 \cdot expand \cdot r^{2}$ parameters. To offset this, MambAdapter shares parameters across adapters: a single pair of projection matrices $(W_{down}, W_{up})$ is reused in all layers. This reduces the parameter cost of the linear projections from $2drl$ to $2dr$, where $l$ is the number of injected adapters. Although this sharing reduces the expressivity of individual layers, the inclusion of Mamba blocks compensates by providing layer-specific modeling capacity.

The pairing of Mamba with bottleneck adapters is theoretically grounded in the compressibility of state-space models. Fundamentally, SSMs compress temporal information into compact hidden-states $h_t \in \mathbb{R}^{N}$ where $N \ll d$. When placed within a bottleneck of dimension $r \ll d$, Mamba operates on a task-specific low-dimensional subspace: the adapter reduces the feature dimension while the SSM further compresses temporal information through its recurrent state updates. Recent work on in-training compression for SSMs demonstrates that their state representations can be substantially reduced with minimal performance degradation, indicating robustness to dimensional constraints \cite{chahine2026curiouscaseintrainingcompression}. Together, these properties suggest that Mamba is architecturally well-matched to low-rank bottleneck adapters, making it particularly suitable for parameter-efficient transfer learning in speech and audio tasks.

%% file: Sections/Experiments.tex
\definecolor{paleturquoise}{HTML}{AFEEEE}

\sethlcolor{green!20}

\section{Experiments}
\label{sec:experiments}

We evaluate the performance and efficiency of our method, MambAdapter, as a PETL technique in both audio/speech classification and multilingual speech recognition settings.

In this work, all adapter experiments adopt parallel adapter insertion\cite{jie2022convolutionalbypassesbettervision, chen2022adaptformeradaptingvisiontransformers, he2022towards, cappellazzo2024efficient}, which we found to yield better performance than sequential insertion\cite{houlsby2019parameterefficienttransferlearningnlp, karimi-mahabadi-etal-2021-parameter} in our internal ablation studies. This finding is consistent with prior works \cite{he2022towards,cappellazzo2024parameterefficienttransferlearningaudio}. For all of our experiments, the complete list of hyperparameters is available with the code\footnote{Code available at: \url{https://github.com/salman-ha/MambAdapter}}.

\subsection{Audio \& Speech Classification}
\subsubsection{Experimental Setup}

We adopt the experimental setup from \cite{cappellazzo2024parameterefficienttransferlearningaudio} to ensure direct comparability with prior PETL methods. Specifically, we evaluate MambAdapter on four standard datasets spanning three tasks, using the AST model as the base architecture:

\begin{enumerate}[leftmargin=*,label=\arabic*.]
  \item \textbf{Audio Classification}: ESC-50 (ESC)~\cite{piczak2015dataset}, UrbanSound8K (US8K)~\cite{Salamon:UrbanSound:ACMMM:14}.
  \item \textbf{Keyword Spotting}: Speech Commands V2 (GSC)~\cite{speechcommandsv2}.
  \item \textbf{Intent Classification}: Fluent Speech Commands (FSC)~\cite{lugosch2019speechmodelpretrainingendtoend}.
\end{enumerate}

\textbf{Audio Spectrogram Transformer (AST)}\cite{gong2021astaudiospectrogramtransformer}. AST is an encoder-only Transformer architecture for audio classification based on the Vision Transformer ~\cite{dosovitskiy2021an}. It splits the log-mel spectrogram into patches, and processes them with standard Transformer layers. For our experiments, we use the AST model pretrained on AudioSet\cite{7952261}.

\textbf{Baselines.} We compare MambAdapter against several PETL baselines as introduced in ~\cite{cappellazzo2024parameterefficienttransferlearningaudio}. We follow their adapter rank settings, using a rank of 16 for both MambAdapter and Bottleneck adapters, and 12 for Conformer. 

\subsubsection{Audio \& Speech Classification Results}
Table \ref{tab:main_audio} reports audio classification results, averaged over 5 random seeds for Bottleneck, Conformer, MambAdapter, and LoRA.

In the \textit{Pfeiffer} adapter configuration, the Conformer adapter achieves the highest average accuracy of 90.07\%, while the Bottleneck adapter lags behind by 8\%. Notably, MambAdapter performs only 0.35\% worse than Conformer, while using less than 25\% of its trainable parameters. Additionally, MambAdapter outperforms Conformer on the US8K dataset by $\sim$ 0.5\%.

Under the \textit{Houlsby} configuration, MambAdapter achieves the highest average accuracy of 89.85\%, surpassing all baselines. Despite using less than 20\% of the parameters of Conformer, MambAdapter delivers competitive or superior performance across most datasets. Its largest improvement is on ESC, where it achieves the highest accuracy overall, even exceeding that of full fine-tuning (FFT).

\begingroup
\setlength{\tabcolsep}{3.3pt}
\begin{table}[t]
\centering
\caption{Accuracies on audio and speech classification tasks. Methods marked with $^*$ are taken from \cite{cappellazzo2024parameterefficienttransferlearningaudio}. Higher is better.}
\label{tab:main_audio}
\begin{tabular}{lcccccc}
\toprule
\textbf{Method} & \textbf{Par (M)} &\cellcolor{pastelyellow} \textbf{ESC} &\cellcolor{lightgray} \textbf{US8K} &\cellcolor{aquamarine} \textbf{GSC} &\cellcolor{melon} \textbf{FSC} & \cellcolor{champagne}\textbf{Avg}\\
\midrule
\textcolor{darkgray}{FFT$^*$} & \textcolor{darkgray}{85} &\textcolor{darkgray}{87.48} &\textcolor{darkgray}{84.31} &\textcolor{darkgray}{97.31} &\textcolor{darkgray}{93.29} &\textcolor{darkgray}{90.07} \\
\hline \addlinespace[2pt]
BitFit$^*$\cite{ben-zaken-etal-2022-bitfit} &0.12 &86.05 &82.17 &85.51 &63.85 &79.40  \\
DPT$^*$ \cite{lester-etal-2021-power} &0.27 &86.52 &83.67 &89.18 &68.60 &81.99  \\
Pref-T$^*$ \cite{li2021prefixtuning} &0.26 &82.93 &81.39 &83.46 &55.75 &75.88  \\ 
LoRA &0.26 &86.45 &81.89 &93.61 &76.00 &84.49\\
\hline
 \rowcolor{pastelviolet}
\multicolumn{7}{c}{\textbf{\textit{Pfeiffer Adapters}}}\\
\hline \addlinespace[2pt]
Bottleneck &0.25 &86.31 &80.69 &90.66 &72.58 &82.56\\\addlinespace[.05em]
Conformer &0.27 &\textbf{87.28} &82.04 &\textbf{94.47} &\textbf{96.49} &\textbf{90.07}\\
\hdashline \addlinespace[1pt]
\textit{MambAdapter} & \textbf{0.06} & \textbf{87.28} & \textbf{82.51} & 94.12 & 94.98 & 89.72\\

\hline
 \rowcolor{lightblue}
\multicolumn{7}{c}{\textbf{\textit{Houlsby Adapters}}}\\
\hline \addlinespace[2pt]

Bottleneck &0.49 &87.15 &80.58 &91.30 &76.67 &83.92 \\\addlinespace[.05em] 
Conformer &0.54 &84.98 &\textbf{82.41} &\textbf{94.91} &\textbf{96.46} & 89.69\\ \hdashline \addlinespace[1pt]
\textit{MambAdapter} & \textbf{0.11} & \textbf{87.55} & 81.71 & 94.27 & 95.85 & \textbf{89.85}\\

\bottomrule
\end{tabular}
\end{table}

\subsection{Automatic Speech Recognition}

\subsubsection{Experimental Setup}
To evaluate the performance of MambAdapter on automatic speech recognition (ASR), we fine-tune Whisper on five languages from Common Voice 13~\cite{ardila-etal-2020-common}. Our experiments are implemented using SpeechBrain~\cite{speechbrain_v1, dellalibera2023clmasrcontinuallearningbenchmark}.

\textbf{Datasets.} The selected target languages are Abkhaz (AB), Central Kurdish (CKB), Esperanto (EO), Kabyle (KAB), and Kinyarwanda (RW), representing low- to medium-resource languages where Whisper's zero-shot performance is relatively weak. Each language's training set contains 600 minutes of transcribed audio, except for Central Kurdish, which includes 484 minutes. All validation and test sets are 60 minutes long.

\textbf{Whisper} \cite{radford2022robustspeechrecognitionlargescale}. Whisper is an encoder–decoder Transformer that serves as a foundation model for speech-related tasks. It was pretrained on $\sim$680,000 hours of multilingual, multitask web-scraped audio, and exhibits strong performance for multilingual automatic speech recognition. It processes log-mel spectrogram features with a convolutional front end, and generates text tokens auto-regressively, guided by task and language-prefix tokens.

\textbf{Baselines.} We compare MambAdapter to LoRA, Bottleneck, Conformer, and full fine-tuning. Due to the observed degradation of performance with lower-rank adapters in ASR, we increase the bottleneck dimension: setting the rank to 32 for LoRA, 64 for Bottleneck, 48 for Conformer, and 104 for MambAdapter in order to match the number of parameters present in the other methods. All adapters use the Pfeiffer configuration and are inserted only in Whisper's encoder, with the decoder remaining frozen.

\subsubsection{Automatic Speech Recognition Results}
Table~\ref{tab:main_asr} reports the word error rate (WER) averaged over 5 seeds. Among the PETL approaches, MambAdapter achieves the best average performance most of the time while maintaining a comparable parameter budget. On average, it reduces WER by 0.8\% compared to Bottleneck adapters and by a more substantial 5.8\% and 7.4\% compared to Conformer and LoRA respectively. Although full fine-tuning remains the strongest overall, adapters can narrow the performance gap to just 4\% WER despite training only 0.45\% of Whisper’s parameters. This highlights the effectiveness of PETL for low- and medium-resource ASR tasks.

\begingroup
\setlength{\tabcolsep}{3.3pt}
\begin{table}[t]
\centering
\caption{Word Error Rates (WER\%) for speech recognition tasks. Lower is better.}
\label{tab:main_asr}
\begin{tabular}{lccllccc}
\toprule
\textbf{Method}
  & \textbf{Par (M)}
  & \cellcolor{pinksecondbest}\textbf{AB}
  & \cellcolor{paleturquoise}\textbf{CKB}
  & \cellcolor{slateblue}\textbf{EO}
  & \cellcolor{teagreen}\textbf{KAB}
  & \cellcolor{lightgray}\textbf{RW}
  & \cellcolor{champagne}\textbf{Avg} \\
\midrule
\textcolor{darkgray}{FFT}            & \textcolor{darkgray}{241}     & \textcolor{darkgray}{46.8}    & \textcolor{darkgray}{46.2}    & \textcolor{darkgray}{18.2}     & \textcolor{darkgray}{53.0}    & \textcolor{darkgray}{61.7}    & \textcolor{darkgray}{45.18}     \\
\hline \addlinespace[2pt]
LoRA & 1.2 & 62.4 & 57.0 & 27.0 & 67.8 & 72.3 & 57.3 \\
Bottleneck        & 1.2    & 54.5  & 50.5 & \textbf{22.2 }   & 60.4 & 65.8 & 50.7  \\
Conformer         & 1.4    & 59.7  & 55.6  & 25.1     & 67.3  & 71.0  & 55.7  \\
\hdashline \addlinespace[1pt]
\textit{MambAdapter} & \textbf{1.1}    & \textbf{53.2}  & \textbf{50.2}  & 22.3    & \textbf{58.8}  & \textbf{64.9}  & \textbf{49.9}   \\
\bottomrule
\end{tabular}
\end{table}
\endgroup

\subsection{Scaling the number of parameters}

To analyze how performance scales with parameter count, we vary the bottleneck dimension \textit{r} and evaluate each method on the GSC dataset using the Pfeiffer configuration.

As shown in Figure~\ref{fig:scaling-gsc}, three key observations emerge. First, Bottleneck adapters consistently underperform compared to both MambAdapter and Conformer, and exhibit minimal performance gains even as their parameter count approaches one million. Second, under constrained parameter budgets (fewer than 500\,k), MambAdapter outperforms both baselines by 0.5--4\%, highlighting its strength in low-resource settings. Third, beyond 600\,k parameters, Conformer surpasses MambAdapter and achieves the highest accuracy, peaking at approximately 95.5\%. However, both MambAdapter and Conformer exhibit diminishing returns beyond this point, suggesting that the majority of performance gains are realized under 500\,k parameters.

\begin{figure}
    \centering
    \includegraphics[width=0.75\linewidth]{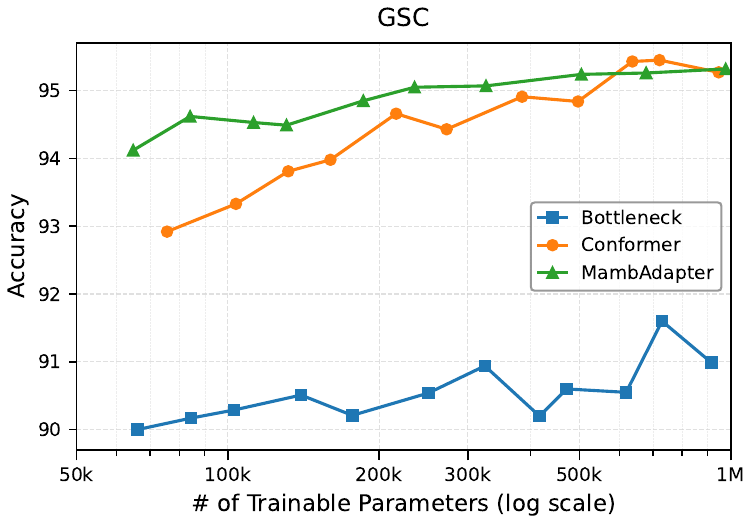}
    \caption{Scaling trend for adapter PETL methods on GSC.}
    \label{fig:scaling-gsc}
\end{figure}

\subsection{Architecture Ablation Study}
We evaluate the impact of key architectural components in MambAdapter by conducting ablation studies across three audio and speech classification datasets: ESC, GSC, and FSC. Specifically, we examine the contribution of (1) the Mamba block, (2) the per-layer scaling factor $\alpha$, and (3) the shared linear projections. The results are summarized in Table~\ref{tab:ablation}.

\textbf{Removing Mamba.}  
When the Mamba block is removed, leaving only the shared bottleneck and the per-layer scaling factor $\alpha$, accuracy drops by 3--4\% on ESC and GSC, and by about 30\% on FSC. Notably, this configuration performs comparably to standard bottleneck adapters on GSC but underperforms them by $\sim$4\% on ESC and $\sim$7\% on FSC. These results highlight the significant capacity contributed by the Mamba blocks, particularly for more complex tasks like intent classification in FSC.

\textbf{Removing the scaling factor $\alpha$.}  
Eliminating $\alpha$ while retaining the Mamba block allows us to assess its contribution as the sole source of per-layer variation. This results in a modest performance drop of less than 2\% across all tasks. Although small, this decrease is consistent, indicating that $\alpha$ provides beneficial per-layer flexibility at negligible cost.

\textbf{Disabling parameter sharing.}  
We replace the shared down- and up-projection matrices with independently learned projections for each adapter layer. This leads to a slight performance improvement on GSC and FSC (1--2\%), but a small decrease on ESC ($\sim$2\%). This suggests that removing projection sharing on ESC may introduce mild over-parameterization, whereas sharing the projections provides a regularizing effect. Compared to MambAdapter, this variant yields an average accuracy gain of less than 1\%, but requires more than four times the number of trainable parameters. This highlights the trade-off between increased capacity and parameter efficiency.

\begin{table}
\centering
\caption{Ablation study for the design of MambAdapter.}
\label{tab:ablation}
\begin{tabular}{lccccc}
\toprule
\textbf{Method}      & \textbf{Par (M)} & \cellcolor{pastelyellow}\textbf{ESC} & \cellcolor{aquamarine}\textbf{GSC} & \cellcolor{melon}\textbf{FSC} & \cellcolor{champagne}\textbf{Avg} \\
\midrule
\textbf{MambAdapter} & {0.06} & {\textbf{87.28}} & {94.12} & {94.98} & {92.12}\\
\hline \addlinespace[2pt]
No Mamba & \textbf{0.04} & 84.32 & 90.40 & 65.38 & 80.03 \\
Not Scaled & 0.06 & 85.45 & 94.02 & 94.71 & 91.39 \\
\hdashline \addlinespace[2pt]
Not Shared & 0.28 & 85.65 & \textbf{94.67} & \textbf{96.47} & \textbf{92.26}\\
\bottomrule
\end{tabular}
\end{table}

\subsection{Mamba Hyperparameter Ablation Study}

\newcommand{\werw}{\linewidth}
\begin{figure}
  \centering
  \captionsetup{justification=centering}
  \includegraphics[width=\werw]{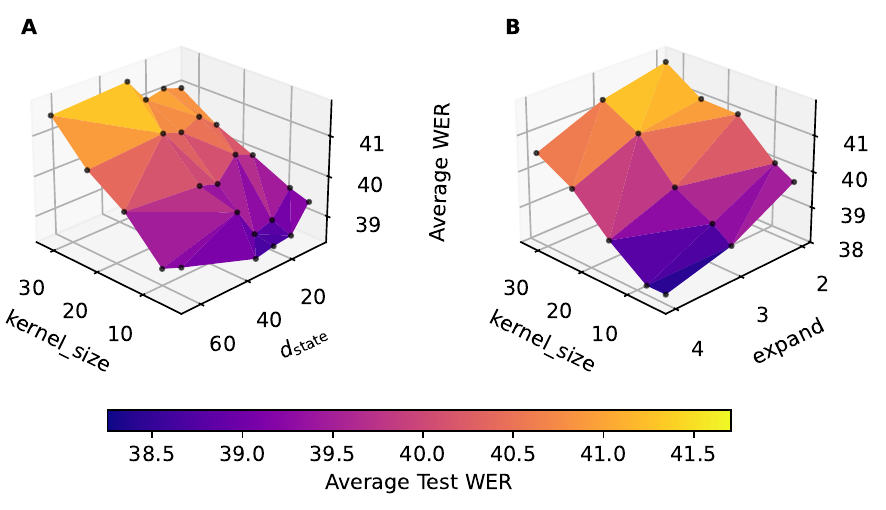}
  \caption{WER $\downarrow$ averaged over \textit{expand} (left) and $d_{\text{state}}$ (right).}
  \label{fig:wer-surface}
\end{figure}

We investigate the impact of Mamba’s hyperparameters—\textit{kernel size}, $d_{\text{state}}$, and \textit{expand}—on ASR performance by evaluating 75 hyperparameter combinations on CKB and EO, then averaging their WER. The results are summarized in Figure~\ref{fig:wer-surface}, which shows the word error rate (WER) averaged across \textit{expand} (left) and $d_{\text{state}}$ (right).

We can observe a few noticeable trends. (1) Increasing the kernel size causes a marginal yet consistent decrease in performance. This can be seen by the steady increase of WER with kernel size on both plots. (2) Increasing $d_{\text{state}}$ does not guarantee a performance gain; instead, an intermediate range ($20$–$40$) yields the best trade-off. This suggests that excessively enlarging the recurrent memory capacity introduces redundancy, making it wasteful for ASR. (3) With every increase in \textit{expand}, we observe a performance gain of roughly $1\%$ WER. This shows that larger \textit{expand} improves performance by widening the intermediate representations for the state update.

\begin{table}
\centering
\caption{Latency (ms) and peak GPU memory (MB).}
\label{tab:computational_cost}
\begin{tabular}{lcc|cc|cc}
\toprule
 & \multicolumn{2}{c}{\textbf{Streaming}} & \multicolumn{2}{c}{\textbf{Mid}} & \multicolumn{2}{c}{\textbf{Batch}} \\
 & \textbf{ms} & \textbf{MB} & \textbf{ms} & \textbf{MB} & \textbf{ms} & \textbf{MB} \\
\midrule
Whisper & 13.03 & 874 & 28.58 & 2199 & 100.66 & 7240 \\
\midrule
w. Bottleneck & 14.04 & 880 & 29.82 & 2206 & 104.44 & 7249 \\
w. Conformer & 16.15 & 884 & 32.07 & 2210 & 111.72 & 7253 \\
w. \textit{MambAdapter} & 17.66 & 881 & 31.62 & 2207 & 110.21 & 7248 \\
\bottomrule
\end{tabular}
\end{table}

\subsection{Latency \& Memory Usage}
We measure inference latency and peak GPU memory when integrating each adapter into Whisper. Measurements are averaged over 30 forward passes on an NVIDIA H100 across three scenarios varying in batch size and audio length: \textit{Streaming} (batch size 1, 5s), \textit{Mid} (8, 10s), and \textit{Batch} (32, 30s), representing real-time, moderate-scale, and offline processing. As shown in Table~\ref{tab:computational_cost}, all adapters exhibit comparable memory usage with a marginal increase relative to Whisper. MambAdapter's latency overhead is largest in \textit{Streaming}, where the fixed costs of the selective scan dominate for short, unbatched inputs, but are amortized with sequence length and batch size. This positions it more for offline and long-form processing than for streaming.

%% file: Sections/Conclusion.tex
\section{Conclusion}
\label{sec:conclusion}

In this work, we introduced MambAdapter, a lightweight adapter design that integrates Mamba blocks with shared projections for parameter-efficient transfer learning. Its consistent strong performance across audio classification and speech recognition makes it a promising option for adapting speech and audio foundation models even under limited budgets.